\documentstyle[aps,multicol,epsf]{revtex}

\begin{document}

\title{Adaptation of Autocatalytic Fluctuations to Diffusive Noise}

\author{N. M. Shnerb$^1$, E. Bettelheim$^2$, Y. Louzoun$^2$, O. Agam$^2$, 
S. Solomon$^2$}

\address{$^1$Department of Physics, Judea and Samaria College, Ariel 44837,
 Israel. }
\address{$^2$Racah institute of Physics, The Hebrew University, Jerusalem 
91910, Israel.}

\date{\today}
\maketitle

\begin{abstract}
Evolution of a system of diffusing and proliferating mortal reactants is analyzed
in the presence of randomly moving catalysts. While the continuum description
of the problem predicts reactant extinction as the average growth rate becomes
negative, growth rate fluctuations induced by the discrete nature of the agents
are shown to allow for an active phase, where reactants proliferate as their
spatial configuration adapts to the fluctuations of the catalysts density. The
model is explored by employing field theoretical techniques, numerical simulations
and strong coupling analysis. For \( d\leq 2 \), the system is shown to exhibits
an active phase at any growth rate, while for \( d>2 \) a kinetic phase transition
is predicted. The applicability of this model as a prototype for a host of phenomena
which exhibit self organization is discussed.
\end{abstract}

\vspace{0.15in}
\begin{multicols}{2}

\section{Introduction}

There is a growing interest in the dynamics of catalytic systems with diffusing
reactants\cite{review}. These models has been considered in the theory of population
biology \cite{Murray}, chemical kinetics \cite{chemical} and physics of contact
process \cite{contact} as well as magnetic systems\cite{grass-magnetic}. In
the most simple case, where agents undergo only birth (autocatalytic reproduction)
and death (spontaneous annihilation), the total growth rate, i.e., the difference
between the typical rates for these two processes, is the critical parameter
for the system. While negative growth rate implies exponential decrease in the
number of particles toward extinction of the ``colony'', positive rate gives
exponential growth. Usually, the number of reactants saturates to some constant
value which reflects the finite holding capacity of the environment. The most
simplified mathematical description of this process is given by the continuum
Fisher equation\cite{fisher}:

\begin{equation}
\label{eq:Gribov}
\frac{\partial b(x,t)}{\partial t}=D\nabla ^{2}b(x,t)+\sigma b(x,t)-\lambda b^{2}(x,t)
\end{equation}
 where \( b(x,t) \) stands for the density of reactants, \( \sigma  \) is
the total growth rate, and \( -\lambda b^{2} \) is introduced phenomenologically
as the ``minimal'' nonlinear term which leads to saturation at positive growth
rate. Since the density \( b \) is positive semi-definite, at negative growth
rate \( (\sigma <0) \) there is only one steady state, the absorbing state,
where \( b(\bf {x},t)=0 \) everywhere. At positive \( \sigma  \), this state
becomes unstable, and the system flows into the uniform state \( \bar{b}=\sigma /\lambda  \).
In this simplified framework the diffusion is irrelevant to the steady state,
and only governs the dynamical approach to it, an effect which has been considered
at \cite{fisher}. It turns out that the typical invasion of the unstable phase
by the stable one is in the form of Fisher fronts (of width \( w\sim \sqrt{D/\sigma } \))
which propagate with velocity \( v\geq 2\sqrt{D\sigma } \). At the stable state,
any small fluctuation with wavelength \( k \) decays as \( \exp [-(\sigma +Dk^{2})t] \).
The phase transition from the inactive to the active state takes place at \( \sigma =0 \). 

Equation (\ref{eq:Gribov}) describes the continuum limit of many underlying
discrete processes. A typical example is a system with particles diffusing (random
walk), annihilating (\( B+B\longrightarrow \emptyset  \)) and reproducing autocatalytically
\( B\longrightarrow 2B \). The discrete nature of individual reactants and
their stochastic motion introduce a (multiplicative) noise, which may dominate
the evolution of the system and violate the predictions of (\ref{eq:Gribov}).
For the above mentioned and similar processes, it turns out that, at low dimensionality
\( (d\leq 2) \), the extinction phase is stable even at small, positive growth
rate, and the transition from active to inactive state falls into the equivalence
class of directed percolation \cite{percolation} (Reggeon field theory). Moreover,
it has been conjectured \cite{grass} that any transition with single absorbing
state falls into the same equivalence class, unless some special symmetry or
conservation laws are introduced, as the even offspring case considered by Grassberger
\cite{grass-magnetic}, and Cardy and Tauber \cite{cardy}. 

Recently, a new type of active-inactive phase transition has been introduced,
for the process (1) in the presence of \textit{quenched} disorder in the relevant
term, \( \sigma (x) \). It has been shown by Janssen \cite{janssen}, that
the renormalization group (RG) flow of that process has only a runaway solutions
in the physical domain (due to the ``ladder'' diagrams which changes the effective
mass of the free propagator), hence the phase transition is \textit{not} of
directed percolation type. Nelson and Shnerb \cite{ns}, using the continuum
approximation, showed that the local growth is related only to \( \sigma (\bf {x}) \)
at the vicinity of the domain, i.e., \textit{localization} of Anderson type
\cite{Anderson} takes place and the extinction transition is given by the effective
growth rate of small, localized islands. Although the effect of intrinsic noise
due to discreteness fluctuations has not been considered in \cite{ns}, it seems
reasonable that the actual transition takes place when the time scale for tunneling
between two positive growth islands is smaller than the time scale for absorbing
state decay of a single ``oasis''. 

If the disorder is \textit{uncorrelated}, \( \sigma (\bf {x},t) \) with \( \left\langle \sigma (\bf {x},t)\sigma (\bf {x'}t')\right\rangle =\Delta \delta (\bf {x-x'})\delta (t-t') \),
the linear part of Eq. (\ref{eq:Gribov}) also governs the statistics of a directed
polymer on a heterogeneous substance, where the time in (\ref{eq:Gribov}) is
identified with the polymer's preferred direction \cite{directed polymer}.
With the Cole-Hopf transformation, this problem is mapped to the noisy Burger's
process \cite{fns} (KPZ surface growth\cite{kpz}). In contrast with the ``localization''
in the case of static disorder, uncorrelated environment induces \textit{superdiffusion}
of the reactants, where the ``center of mass'' of the population wanders in
space as \( r\propto t^{\zeta } \), with \( \zeta >0.5. \) The effect of intrinsic
stochasticity due to discretization is, again, limited, and the statistical
properties of the eigenenergies of the directed polymer problem determines the
extinction transition. This is also the case when the system under quenched
disorder is subject to strong convection, as has been shown by \cite{ns}. 

In this paper, we consider the case of \textit{diffusive} disorder.

\[
B\stackrel{\mu }{\longrightarrow }\emptyset \]

\[
B+A\stackrel{\lambda }{\longrightarrow }2B+A\]
 when both \( B \) and \( A \) undergo diffusion with rates \( D_{b} \) and
\( D_{a} \), respectively. The mortal agent, \( B \), dies at rate \( \mu , \)
and proliferates in the presence of the (eternal) catalyst \( A \). The continuum
description for this process is given by the ``mean-field'' (rate) equations
for the densities \( a(x,t) \) and \( b(x,t) \):

\[
\frac{\partial b(x,t)}{\partial t}=D_{b}\nabla ^{2}b(x,t)-\mu b+\lambda ab\]

\begin{equation}
\label{eq:continuum}
\frac{\partial a(x,t)}{\partial t}=D_{a}\nabla ^{2}a(x,t).
\end{equation}
 As \( t\to \infty , \) \( a \) flows into its average \( \bar{a} \), thus
the effective mortality rate for \( b \) is given by \( m=\mu -\lambda \bar{a} \)
(the mortality rate turns out to be the ``mass'' of the effective field theory,
hence denoted by \( m \)). For positive ``mass'' the \( b \) population
decays exponentially while negative mass implies exponential growth. The active-inactive
phase transition takes place at \( m=0. \) One observes that equation (\ref{eq:continuum})
is obtained from (\ref{eq:Gribov}) by dropping the non-linear term and replacing
\( \sigma  \) by \( \lambda \bar{a}-\mu  \), accordingly, at long times the
process introduced in (\ref{eq:continuum}) is the linearized form of (\ref{eq:Gribov})
with the proliferation rate, \( \sigma  \), fluctuates ``diffusively'' around
its mean \( \lambda \bar{a}-\mu  \). As for this system the disorder is not
static but is correlated, it somehow interpolates between the above mentioned
models and one may wonder weather it leads to localization of the reactants
or to superdiffusion \cite{hwa}. It turns out that the reactants may adapt
themselves to the environment and the colonies are localized on the diffusing
islands. Moreover, these correlated fluctuations due to the stochastic motion
of individual reactants will change the character of the transition; the transition
point is pushed to negative values of \( m \) and the \( b \) reactants survive
below the ``classical'' threshold. Some of our results, along with a numerical
study of the transition, are summarized in previous publication \cite{pnas}.

\section{Strong Coupling Analysis}

The basic intuition beneath the phenomenon we describe is in the concept of
\textit{adaptive fluctuations}. Let us take a look, first, at the case of frozen
\( A \)-s, where we have random, quenched, growth rate as in \cite{ns}. The
linearized continuum equation then takes the form

\[
\frac{\partial b}{\partial t}=D\nabla ^{2}b-m(x)b\]
 where \( m(x)=\mu -\lambda n_{A}(x) \), and \( n_{A}(x) \) is the random
concentration of the catalyst \( A. \) The system is in its active phase if
the linearized evolution operator 

\[
L=D\nabla ^{2}-m(x)\]
 admits at least one positive eigenvalue, and its localization properties are
almost determined by the corresponding eigenfunctions. Since the same operator
governs the physics of quantum particles in random potential, one may use the
known results \cite{Efros} of this field, i.e., that in low dimensionality
or strong disorder all the wavefunctions are exponentially localized and the
diffusion becomes irrelevant on large length scales\cite{tailstates}. Accordingly,
the system may be in its active phase at localized islands even if the average
\( m \) is positive.

In our case, however, the catalysts diffuse, and these colonies survives only
if the reactants cluster is able to ``trace'' a specific catalyst or to find
some other wandering island. This implies the significance of the system dimensionality:
while two typical random walkers (such as the catalyst and the reactant) encounter
each other in finite time for \( d<2 \), they will (typically) never collide
for \( d>2 \). One may expect, accordingly, that below \( 2d \) quantization
induced fluctuations are much more dominant than above two dimensions. 

Consider one ``frozen'' catalyst at the origin. The effective growth rate in
the vicinity of the origin is positive, i.e., \( m(r<R)=m_{in}<0 \) in a region
of typical catalyst size \( R \) around it. In the ``desert'', out of this
island, the ``mass'', \( m_{out} \), is positive. The colony is then localized
at \( r=0 \), with growth rate \( \left| m_{in}\right|  \) and an exponentially
decreasing tail into the desert\cite{ns}. In the continuum approximation, the
time dependent profile of the tail is given by \cite{remark-cutoff}:

\begin{equation}
\label{profile}
b(r)\sim e^{\left| m_{in}\right| t-r\sqrt{m_{out}/D}},
\end{equation}
and the tail front, which is the size of the reactants colony, moves away from
the origin with typical velocity \( v\sim m_{in}\sqrt{D/m_{out}} \). If the
catalyst is moving, the colony will die only when the \( A \) molecule detaches
from the \( B \) colony. This, however is almost impossible for a diffusively
moving \( A \) since the colony's front moves ballistically. As this argument
involves only one catalyst it is independent of system dimensionality. Thus,
at strong

{\narrowtext
\begin{figure}[!t]
\begin{center}
\leavevmode
\epsfxsize=6.0cm
\epsfbox{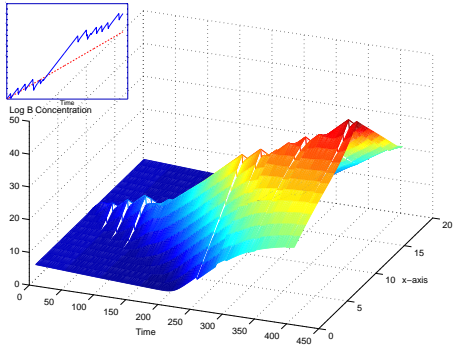} 
\end{center}
\caption{The profile of a B island as a function of time as it follows the random motion
of an A agent. The cross-section of the island is taken through the current
location of the A agent. The inset shows the time evolution of the logarithm
of the height of the B concentration at the point at which A is currently located
(solid line). The B colony is seen to grow, although the average growth rate
over the entire space is negative (\protect\( n_{A}\protect \) is extremely
low since there is only one in the whole simulation space, thus \protect\( \lambda n_{A}-\mu \approx -\mu \protect \)).
The dashed line shows the exponential growth according to Eq. (\ref{eq:expgrowth}).}
\label{singleA}
\end{figure}
}
\noindent 
coupling some localized islands are active in any dimension, in contradiction
with the mean field prediction of extinction at positive average mass. This
mechanism is illustrated in figure (\ref{singleA}), which manifests the ability
of the \( B \) colony to adapt to the location of the moving catalyst. We stress
that this adaptive skill is solely due to the ``dumb'' diffusion and multiplication
of the reactants, and thus is an emergent self-organized feature.

There is, however, a possibility for a different scenario, the weak coupling
limit, where the local properties of the system do not support the formation
of colonies in the inactive phase. For a \( B \) molecule having a spatial
overlap with an \( A \) catalyst, the multiplication time is proportional to
\( \lambda  \). If this time is much larger than the relevant hopping rate,
the typical birth event is singular and no colony is formed. If, furthermore,
the decay time for a \( B \) particle in the absence of \( A \) is much smaller
than the typical time to find a new catalysts one may expect that the system
is in its inactive phase, unless some global, collective effect turns this local
analysis void.

Before looking for global effects, let us try to consider the strong coupling
limit more carefully. Consider a single \( A \) agent located at the point
\( r_{0} \), and the ``island'' of \( B \) reactants which surrounds it.
Keeping the \( A \) stationary and working on a \( d \)-dimensional lattice
(with lattice constant \( l \) , the growth rate is \( \frac{\lambda }{l^{d}} \)
and the hopping rate is \( \frac{2Dd}{l^{2}} \)) the following equation holds
for the concentration of \( B \)'s:

\begin{equation}
\label{eq:stationary}
n_{B}\left( r,t\right) \sim e^{\left( \frac{\lambda }{l^{d}}-\frac{2dD_{B}}{l^{2}}-\mu \right) t}e^{\log \left( \frac{D_{B}}{\lambda }l^{d-2}\right) \frac{\left| r-r_{0}\right| }{l}},
\end{equation}
where we assume \( D_{B}l^{d-2}\ll \lambda  \) (thus the very steep slope suppresses
the effect of diffusion returning inwards). Consequently,

\begin{equation}
\label{eq:raten}
\frac{\partial \log \left[ n_{B}\left( r_{0},t\right) \right] }{\partial t}\sim \left( \frac{\lambda -2dD_{B}}{l^{2}}-\mu \right) .
\end{equation}
 Now consider a hopping event of the A, by a single lattice spacing. Measuring
\( n_{B}\left( r_{0}\left( t\right) ,t\right)  \), at the new \( A \) site,
it reduces by a factor of 
\begin{equation}
\label{eq:reduction}
e^{-\log \left( \frac{D_{B}}{\lambda }l^{d-2}\right) }.
\end{equation}
The rate at which the hopping events occur is \( \frac{2dD_{A}}{l^{2}} \),
accordingly the rate equation (\ref{eq:raten}) is modified to:
\begin{eqnarray}
\frac{\partial \log \left[ n_{B}\left( r_{0}\left( t\right) ,t\right) \right] }{\partial t} &\sim& \nonumber \\
 \frac{\lambda -2dD_{B}}{l^{2}}-&\mu& -\frac{2dD_{A}}{l^{2}}\log \left( l^{d-2}\frac{D_{B}}{\lambda }\right),
\label{eq:lograte}
\end{eqnarray}
where we have assumed that there is enough time between hopping events so that
the island shape stabilizes to the long time behavior (\ref{eq:stationary}),
namely \( D_{A}l^{d-2}\ll \lambda  \). Equation (\ref{eq:lograte}) has the
solution: 
\begin{equation}
\label{eq:expgrowth}
n_{B}\left( r_{0}\left( t\right) ,t\right) \sim e^{\left( \frac{\lambda -2dD_{B}}{l^{2}}-\mu -\frac{2dD_{A}}{l^{2}}\log \left( l^{d-2}\frac{D_{b}}{\lambda }\right) \right) t}.
\end{equation}
This shows that in the strong coupling regime the exponent is positive and the
number of reactants grows exponentially, independent of the dimension and of
the catalyst density. The inset of Fig. (\ref{singleA}) shows the fit of the
expression (\ref{eq:expgrowth}) to the numerical results on a lattice.

\section{Weak Coupling}

In order to consider global effects of spatial fluctuations, let us write the
Master equation for the probability \( P_{nm} \) to find \( m \) reactants
and \( n \) catalysts at a single point (with no diffusion)

\begin{eqnarray}
\frac{dP_{nm}}{dt}=-&\mu& [mP_{nm}-(m+1)P_{n,m+1}]\nonumber \\
&-&\lambda [mnP_{nm}-n(m-1)P_{n,m-1}].
\end{eqnarray}

Following \cite{Doi-Pelilti} we define a set of creation-annihilation operators,

\[
\begin{array}{cc}
a^{+}\left| n,m\right\rangle =\left| n+1,m\right\rangle  & b^{+}\left| n,m\right\rangle =\left| n,m+1\right\rangle \\
a\left| n,m\right\rangle =n\left| n-1,m\right\rangle  & b\left| n,m\right\rangle =m\left| n,m-1\right\rangle 
\end{array}\]
 and a wavefunction

\[
\Psi =\sum _{n,m}P_{n,m}\left| n,m\right\rangle .\]
 The Master equation then takes the Hamiltonian form:

\[
\frac{\partial \Psi }{\partial t}=-\mbox {H}\Psi \]
 with

\begin{eqnarray}
\mbox {H}&=& \sum_{i} \left[ \frac{D_{a}}{l^{2}}\sum _{<e-i>}a^{+}_{i}\left
( a_{i}-a_{e}\right) +\frac{D_{b}}{l^{2}}\sum _{<e-i>}b^{+}_{i}\left( b_{i}-
b_{e}\right) \right. \nonumber \\ 
 &+& \left. \mu [b_{i}^{+}b_{i}-b_{i}]+\frac{\lambda }{l^{d}}[a_{i}^{+}a_{i}b_{i}^{+}
b_{i}-a_{i}^{+}a_{i}b_{i}^{+}b_{i}^{+}b_{i}] \right]
\end{eqnarray}
 Where \( i \) runs over all lattice points, and the sum, $\left<i-e \right>$,
 is
over nearest neighbors. 

Shifting the creation operators to their vacuum expectation value \( a^{+}\rightarrow \overline{a}+1 \)
and \( b^{+}\rightarrow \overline{b}+1 \), the value of the catalyst density
to its average, \( \frac{1}{l^{2}}a\rightarrow a+n_{a}, \) and finally, \( \frac{1}{l^{2}}b\rightarrow b \)
the evolution operator takes the following form in the continuum limit:

\begin{eqnarray}
\mbox {H}=\int d^{d}x &[& -D_{b}\overline{b}\nabla ^{2}b-D_{a}\overline{a}\nabla ^{2}a+\mu \overline{b}b \nonumber \\ 
&-&\lambda \overline{b}b(\overline{a}+1)(\overline{b}+1)(a+n_{a})]
\end{eqnarray}
 The action is simply:

\begin{equation}
\label{eq:action}
S=\int dt\left[ \int d^{d}x\left( \overline{a}\frac{\partial }{\partial t}a+\overline{b}\frac{\partial }{\partial t}b\right) +\mbox {H}\right] .
\end{equation}
Note that the coefficient of \( \overline{b}b \), which plays the role of mass,
is given by \( m\equiv \mu -\lambda n_{a} \). 

Now this system may be analyzed using the standard renormalization group (RG)
technique. We impose a change of scale

\begin{equation}
\label{rescale}
\begin{array}{c}
x\to sx\\
t\to s^{z}t\\
a\to s^{-d-\eta }a\\
\Lambda \to \Lambda /s
\end{array}
\end{equation}
 where \( s \) is the renormalization group scale factor. The renormalization
flows of the parameters of the action (\ref{eq:action}) are given by their
naive dimensionality and the corrections from the diagrams shown in Fig. (\pageref{fig:renormalizer}),
using the basic vertices as in Fig. (\ref{fig:vertices}). The flow equations
for the mass and the coupling constant are given by

\[
\frac{d\lambda }{d\, \ln (s)}=\epsilon \lambda +\frac{\lambda ^{2}}{2\pi D}\frac{\Lambda ^{d-2}}{1+\frac{m}{D\Lambda ^{2}}}\]

\begin{equation}
\label{RG}
\frac{dm}{d\, \ln (s)}=2m-\frac{\lambda ^{2}n_{a}}{2\pi D}\frac{\Lambda ^{d-2}}{1+\frac{m}{D\Lambda ^{2}}},
\end{equation}
 where \( \epsilon =2-d, \) \( D=\frac{D_{a}+D_{b}}{2} \) and \( \Lambda (s) \)
is the upper momentum cutoff. Note that, as indicated by naive dimensional analysis,
the Gaussian fixed point \( \{\lambda =0,m=0\} \) is stable at \( d>2, \)
hence there is no perturbative corrections to \( \eta  \) and \( z \) in this
regime. If the momentum cutoff is much larger than any other quantity of the
problem one has,

\[
\frac{d\lambda }{d\, \ln (s)}=\lambda \left( \epsilon +\frac{\lambda }{2\pi D}\right) \]
 
\[
\frac{dm}{d\, \ln (s)}=2m-\frac{\lambda ^{2}n_{0}}{2\pi D}\]
 and the flow lines are shown in Figs. (\ref{dl2}) and (\ref{fig:dg2}) for
\( d\leq 2 \) and \( d>2 \) , respectively. Below two dimensions, the Gaussian
fixed point is always unstable and the system flows to the strong coupling limit,
where adaptive B colonies grow indefinitely, as indicated by the negative values
of the effective mass. At higher dimensionality, on the other hand, there is
a finite region in the parameter space where the trivial fixed point is stable
and \( \lambda  \) flows into zero, while higher values of initial coupling
constant flow to infinity. For a system of finite size \( L^{d} \), the flows
should be truncated at \( s=\frac{L}{l}, \) and the phase is determined by
the end point of the flow lines at this \( s. \) For finite \( \Lambda  \),
equations (\ref{RG}) take the form:

\[
\frac{d\gamma }{d\, \ln (s)}=\epsilon \gamma +\frac{\gamma ^{2}}{1+M}\]

\[
\frac{dM}{d\, \ln (s)}=2M-\frac{\alpha \gamma ^{2}}{1+M},\]
 where the dimensionless quantities are \( \gamma =\frac{\lambda \Lambda ^{d-2}}{2\pi D} \),
\( M=\frac{m}{D\Lambda ^{2}}, \) and \( \alpha =\frac{2\pi n_{a}}{\Lambda ^{d}}. \)
The flow lines for \( d=2 \) are shown in figure (\ref{dl2new}) and exhibit
a transition due to the finite lattice spacing.

\vspace{0.3cm}

{\narrowtext
\begin{figure}[h]
\vspace{-0.15in}
\begin{center}
\leavevmode
\epsfxsize=7.0cm
\epsfbox{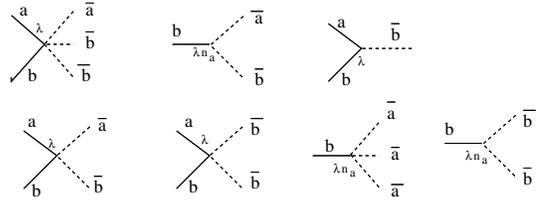} 
\end{center}
\caption{Elements of perturbative expansion: the vertices corresponding
to the action (\ref{eq:action})}
\label{fig:vertices}
\vspace{-0.26cm}
\end{figure}
}

{\narrowtext
\begin{figure}[h]
\vspace{-0.15in}
\begin{center}
\leavevmode
\epsfxsize=4.0cm
\epsfbox{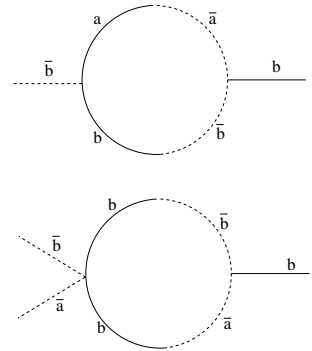} 
\end{center}
\caption{ Most UV divergent diagrams contributing to the renormalization
group equations.}
\label{fig:renormalizer}
\vspace{-0.26cm}
\end{figure}
}

{\narrowtext
\begin{figure}[h]
\vspace{-0.15in}
\begin{center}
\leavevmode
\epsfxsize=6.0cm
\epsfbox{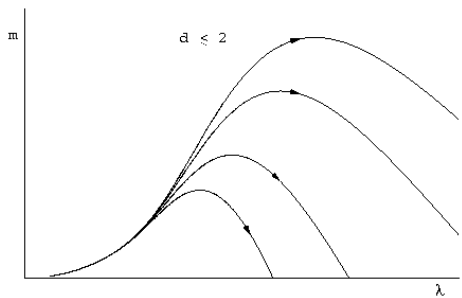} 
\end{center}
\caption{ The RG flow lines at the continuum limit for \protect\( d\leq 2.\protect \)
While at short times \protect\( m\protect \) grows, the system flows into its
active phase \protect\( m<0\protect \) on large time scales.}
\label{dl2}
\vspace{-0.26cm}
\end{figure}
}

{\narrowtext
\begin{figure}[h]
\vspace{-0.15in}
\begin{center}
\leavevmode
\epsfxsize=6.0cm
\epsfbox{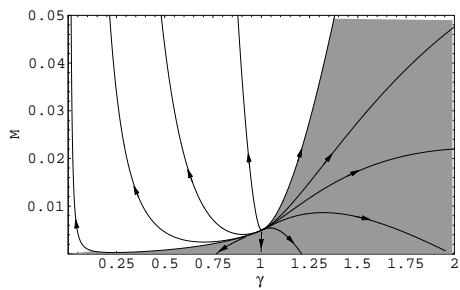} 
\end{center}
\caption{Renormalization flows for \( d>2 \). The shaded region flows into to active
phase while at the unshaded region the system flows to the inactive phase (\( m\to \infty ). \)}
\label{fig:dg2}
\vspace{-0.26cm}
\end{figure}
}

\vspace{0.3cm}

{\narrowtext
\begin{figure}[h]
\vspace{-0.15in}
\begin{center}
\leavevmode
\epsfxsize=6.0cm
\epsfbox{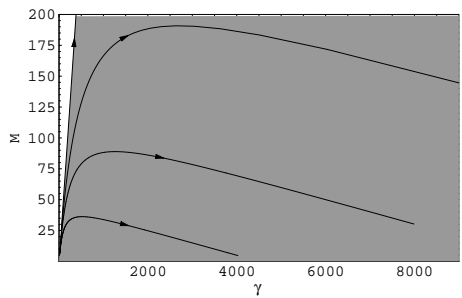} 
\end{center}
\caption{Renormalization flows for \( d=2 \) at finite lattice spacing. Unlike (\ref{dl2}),
here there is a region in parameter space (unshaded) where the extinction phase
is stable.}
\label{dl2new}
\vspace{-0.26cm}
\end{figure}
}

\section{Discussion}

Since the classical works of Malthus and Verhulst\cite{Murray}, it has been
recognized that most of the processes in living systems are autocatalytic and
thus are characterized by exponential growth. In fact, the appearance of an
autocatalytic molecule may be considered as the origin of life. In this paper,
these autocatalytic system are shown to admit self organization in the presence
of fluctuating environment. The exponential amplification of ``good'' fluctuations
in the catalysis parameters prevails, in the situations discussed above, the
globally hostile environment and is robust against the random motion of both
the reactants and the catalysts. Our result may be interpreted as an indication
that ``life'' (in the above sense) is resilient and is able to adapt itself
to the changing environment. The applicability of this model ranges from biological
evolution (where the environment is the genome space) to the role of enzymes
in chemical reactions and even in social or financial settings.

More realistic models, however, should take into account the depletion of resources
by the catalytic process and the finite carrying capacity of the substrate.
Although the model discussed above is relevant at time scales which are small
in comparison with the mean time for consumption or saturation, the stable fixed
point of the system may be different. In particular, on a uniform, inexhaustible,
autocatalytic substrate with finite carrying capacity the discreteness induced
fluctuations have been shown \cite{cardy} to \textit{decrease} the effective
growth rate, and to give a directed percolation type transition at \( d<2. \)
The competition between this effect and the effect of adaptive fluctuations
will be considered elsewhere.

\section{Acknowledgements}

We wish to thank P. W. Anderson, P. Grassberger, D. R. Nelson and D. Mukamel
for helpful discussions and comments. O.A. thanks the support of the Israeli
Science Foundation founded by the Israeli Academy of Science and Humanities
and by Grant No. 9800065 from the USA-Israel Binational Science Foundation (BSF).

\end{multicols}

\begin{thebibliography}{}
\bibitem{remark-cutoff}We may approximate the effect of reactant quantization by taking a lower cutoff
for \( b(r) \), e.g., \( b(r) \) is given by (\ref{profile}) only for \( b>const \)
and is zero elsewhere.
\end{thebibliography}
\end{document}